\documentclass[aps,pra,preprint,amsmath,amssymb,showpacs]{revtex4}
\usepackage{graphicx}
\usepackage{amsmath}
\usepackage{dcolumn}
\usepackage{bm}

\newcommand{\bq}{\mathbf{q}}
\newcommand{\bk}{\mathbf{k}}
\newcommand{\br}{\mathbf{r}}

\newcommand{\eff}{\mathrm{eff}}

\newcommand{\ha}{\hat{ a} }

\newcommand{\ra}{\rangle}
\newcommand{\la}{\langle}

\newcommand{\intek}{ \int \tilde{d} \, \bk}

\newcommand{\tek}{  \tilde{d} \, \bk}

\newcommand{\teq}{  \tilde{d} \, \bq}

\newcommand{\vepsilon}{\bm \epsilon}

\newcommand{\vx}{\bm x}
\newcommand{\vy}{\bm y}
\newcommand{\vz}{\bm z}

\newcommand{\ve}{\bm e}

\begin{document}
\title{Notes on Polarization Measurements}
\author{A. Aiello}
\author{J.P. Woerdman}
\affiliation{Huygens Laboratory, Leiden University\\
P.O.\ Box 9504, 2300 RA Leiden, The Netherlands}
\begin{abstract} We review a few useful concepts
about polarization measurements in the quantum domain. Using a
perfectly general formalism, we show how to build the quantum
counterpart of some classical quantities like  Stokes parameters
and  Mueller matrices, which are well known in classical
polarization-measurement theory.
\end{abstract}
 \maketitle
%
%
%
%
%

\section*{The effective polarization density matrix}
In these notes, our first aim is to re-derive the formulas for the
effective density matrix $\rho_\eff$ we introduced in
\cite{Aiello04c}, without using the concept of Stokes parameters
and without using a circular polarization basis.
To begin with, let us establish our notation. In order to make the
results comparable with the ones in Refs. \cite{Peres_et_al}, we
adopt a relativistic notation. All formulas are given in natural
units $(c = \hbar = 1)$ and all quantized fields are transverse.
In this context, single-photon plane-wave states are denoted by
$|\bk, \lambda \ra $ where $\bk$ is the spatial part of the four
momentum $k = (k^0, \bk), \; k^0 = |\bk| \equiv \omega$, and
$\lambda = 1,2$ is the {\em linear} polarization. They are created
from the vacuum state $| 0 \ra$ by the corresponding creation
operator $\ha_{\lambda}^\dagger(\bk )$ which, together with the
annihilation operator $\ha_{\lambda}(\bk )$, satisfies the
commutation relation
\begin{equation}\label{10} \bigl[ \ha_{\lambda}(\bk ),
\ha^\dagger_{\lambda'}(\bk' ) \bigr] = (2 \pi)^3 2 k^0
\delta_{\lambda \lambda'} \delta^{(3)}(\bk - \bk').
\end{equation}
With this convention, the normalization condition between two
plane-wave states reads
\begin{equation}\label{20} \la \bk, \lambda | \bk', \lambda'
\ra = (2 \pi)^3 2 k^0 \delta_{\lambda \lambda'} \delta^{(3)}(\bk -
\bk'),
\end{equation}
which can be obtained directly  by calculating the expectation
value with respect to the vacuum state of the left side of Eq.
(\ref{10}). Then the resolution of the identity can be written as
\begin{equation}\label{30}
 1 = |0\ra \la 0| +  \int \tilde{d} \, \bk \sum_{\lambda = 1}^2 |\bk, \lambda \ra \la \bk, \lambda
 |+ \sum \left\{ \mathrm{multiparticle} \; \mathrm{states}  \right\},
\end{equation}
where $\tilde{d} \, \bk$ denotes the Lorentz-invariant measure
\begin{equation}\label{40} \tilde{d} \, \bk = \frac{d^3
\bk}{(2 \pi)^3} \frac{1}{2 k^0}.
\end{equation}
The plane-wave  states $| \bk, \lambda \ra$ are quite special
since they are eigenstates of the linear momentum of the field.
More generally, a single-photon state can be described by the
density operator
\begin{equation}\label{45}
\hat{\rho} = \sum_{\lambda, \lambda'}^{1,2} \intek \, \tek'
\rho_{\lambda \lambda'}(\bk, \bk') | \bk, \lambda \ra \la \bk',
\lambda' |,
\end{equation}
where $\rho_{\lambda \lambda'}(\bk, \bk')$ is a Hermitian positive
semidefinite $2 \times 2$ matrix. While the linear momentum and
the polarization degrees of freedom are decoupled for photons in
the plane-wave states $|\bk, \lambda \ra $ \cite{Aiello04c}, this
is not the case for photons described by $\hat{\rho}$ which spans
an $\infty^2$-dimensional vector space ($2$ polarization
directions for each wave vector $\bk$). Therefore, if one want to
deal with polarization degrees of freedom only, it becomes
necessary to { reduce} $\hat{\rho}$ to an {\em effective}
finite-dimensional representation $\rho_{\mathrm{eff}}$ ($2 \times
2$ or $3 \times 3 $) \cite{Peres_et_al}. Now we present a
derivation of $\rho_\mathrm{eff}$ which enlightens the geometrical
aspects of the problem.
To this end, we have to do some manipulations on the fields. In
the Coulomb gauge, which is assumed to hold throughout this notes,
the free field vector potential operator $\hat{\mathbf{A}}(\br,t)$
is written
\begin{equation}\label{50} \hat{\mathbf{A}}(\br,t) =  \int
\tilde{d} \, \bk  \sum_{\lambda = 1}^2\bigl[
\vepsilon^{(\lambda)}(\bk)\ha_{\lambda}(\bk ) e^{-i (k^0 t - \bk
\cdot \br)} + \mathrm{h. c.} \bigr],
\end{equation}
 where the real unit vectors $\vepsilon^{(\lambda)} (\bk)$ ({\em linear} polarization) satisfy the
following orthogonality conditions:
\begin{equation}\label{60} \vepsilon^{(1)}(\bk) \cdot
\vepsilon^{(2)}(\bk) = 0, \qquad \vepsilon^{(1)}(\bk) \times
\vepsilon^{(2)}(\bk) = \frac{\bk}{k^0}.
\end{equation}
Since we are working with the {\em transverse} field ($A^0(\br, t)
= 0$), we can assume an Euclidean metric in the spatial
3-dimensional space and make no distinction between high and low
indices.
 It is useful to define $\vepsilon^{(3)}(\bk) \equiv
{\bk}/{k^0}$ and to build the $\bk$-dependent complete basis
$\mathcal{E}(\bk)= \{ \vepsilon^{(1)}(\bk), \vepsilon^{(2)}(\bk),
\vepsilon^{(3)}(\bk) \}$.
Let us consider now three unit vectors $\vx, \vy, \vz$ of a
Cartesian coordinate system or, more generally, a set
$\mathcal{R}$ of three real orthogonal unit vectors $\mathcal{R} =
\{ \ve^{(1)}, \ve^{(2)}, \ve^{(3)} \}$:
\begin{equation}\label{70} \ve^{(a)}\cdot \ve^{(b)}
= \delta^{ab}, \qquad (a,b =1, \ldots, 3),
\end{equation}
which form a complete basis in the ordinary Euclidean
3-dimensional space:
\begin{equation}\label{80} \sum_{a =1}^3 \ve^{(a)} : \ve^{(a)} = \mathbf{1} \quad
\Leftrightarrow \quad\sum_{a =1}^3 e_i^{(a)} e_j^{(a)} =
\delta_{ij}, \qquad (i,j =1, \ldots, 3),
\end{equation}
where the unit dyadic $\mathbf{1}$ has been written as a sum of
dyadic products $\ve^{(a)}: \ve^{(a)}$ ($a = 1,2,3$).
The vectors $\ve^{(a)}$ define three orthogonal {\em spatial}
orientations and they are independent from the {\em momentum}
direction $\bk/k^0$. However, for a given $\bk$, one can write the
orthogonal transformation $\Lambda(\bk)$ between the two basis
$\mathcal{E}(\bk)$ and $\mathcal{R}$ as
\begin{equation}\label{90}
\begin{array}{rcl}
\Lambda_{ab}(\bk) & \equiv & \ve^{(a)} \cdot \vepsilon^{(b)}(\bk)\\
& = & \displaystyle{\sum_{i = 1}^3 e^{(a)}_i
\epsilon^{(b)}_i(\bk)}.
\end{array}
\end{equation}
Then we can write
\begin{equation}\label{100} \vepsilon^{(\lambda)} (\bk) = \sum_{b
=1}^3 \ve^{(b)} \Lambda_{b \lambda} (\bk), \qquad (\lambda =
1,2,3),
\end{equation}
and insert this formula in Eq. (\ref{50}) in order to obtain:
\begin{equation}\label{110}
\hat{\mathbf{A}}(\br,t)= \sum_{b =1}^3 \ve^{(b)} \hat{A}_b(\br,t),
\end{equation}
where we have defined the $b$-th component of the field
$\hat{\mathbf{A}}(\br,t)$ as
\begin{equation}\label{120}
\begin{array}{rcl}
\hat{A}_b(\br,t) & = & \displaystyle{  \int \tilde{d} \, \bk
\sum_{\lambda = 1}^2\bigl[ \Lambda_{b \lambda} (\bk)
\ha_{\lambda}(\bk ) e^{-i (k^0 t - \bk
\cdot \br)} + \mathrm{h. c.} \bigr]}\\\\
& \equiv & \displaystyle{\int \tilde{d} \, \bk \,
\hat{\mathcal{A}}_b (\bk) e^{-i (k^0 t - \bk \cdot \br)} +
\mathrm{h. c.}} ,
\end{array}
\end{equation}
where we have defined the  transformed annihilation  operators
$\hat{\mathcal{A}}_b(\bk )$ ($b = 1,2,3$), as:
\begin{equation}\label{130}
\hat{\mathcal{A}}_b(\bk ) \equiv \sum_{\lambda = 1}^2  \Lambda_b
\/_\lambda (\bk) \ha_{\lambda}(\bk ), \qquad (b=1,2,3).
\end{equation}
It is easy to check that these operators satisfy the following
commutation relations
\begin{equation}\label{140}
\bigl[\hat{\mathcal{A}}_a(\bk),
\hat{\mathcal{A}}_{a'}^\dagger(\bk') \bigr] = (2 \pi)^3 2 k^0
\Delta_{a a'} \delta^{(3)} (\bk - \bk'),
\end{equation}
where we have defined the transverse  Kronecker symbol $\Delta_{a
a'}$ as
\begin{equation}\label{150}
\Delta_{aa'} \equiv \delta_{aa'} - \frac{k_a k_{a'}}{|\bk|^2}.
\end{equation}
As expected, the longitudinal part $- \frac{k_a k_{a'}}{|\bk|^2}$
of $\Delta_{aa'}$, spoils the   {\em canonical} commutation
relation.

In the quantum theory of photo-detection it is a standard practice
\cite{MandelBook} to define the positive frequency operators
$\hat{V}_b(\br,t)$  ($b = 1,2,3$)  as
\begin{equation}\label{160}
\hat{V}_b(\br,t) \equiv \intek \sqrt{2 k^0} \hat{\mathcal{A}}_b
(\bk) e^{-i (k^0t - \bk \cdot \br)},
\end{equation}
and such that $\sum_{b=1}^3 \hat{V}_b^\dagger(\br,t)\hat{V}_b
(\br,t)$ represent the photon density in $(\br,t)$. These
operators can be used to build the {\em polarization correlation}
operators
\begin{equation}\label{170}
\begin{array}{rcl}
  \hat{J}_{ab} & \equiv & \displaystyle{\int d^3 \br \, \hat{V}_a^{ \dagger} (\br,t) \hat{V}_b}(\br,t) \\\\
   & = & \displaystyle{ \intek \, \hat{\mathcal{A}}_a^{ \dagger}(\bk)
   \hat{\mathcal{A}}_b(\bk)},
\end{array}
\end{equation}
where the last result follows immediately from Eq. (\ref{160}).
The meaning of the matrix operator $\hat{J} \equiv
||\hat{J}_{ab}||$ becomes clear when we calculate its trace:
\begin{equation}\label{180}
\begin{array}{rcl}
  \mathrm{Tr} \hat{J} & = & \displaystyle{\sum_{a = 1}^3  \hat{J}_{aa} } \\\\
   & = &  \displaystyle{}\intek \sum_{\lambda =1}^2
   \ha_\lambda^\dagger (\bk) \ha_\lambda(\bk) \equiv  \hat{N},
\end{array}
\end{equation}
where  $\hat{N}$ denotes the photon-number operator.

Now, by inserting the identity resolution Eq. (\ref{30}) in Eq.
(\ref{170}), we obtain
\begin{equation}\label{190}
\begin{array}{rcl}
  \hat{J}_{ab}  & = & \displaystyle{\intek \, \hat{\mathcal{A}}_a^{
  \dagger}| 0 \ra \la 0 |
   \hat{\mathcal{A}}_b} + \sum \left\{ \mathrm{multiparticle} \; \mathrm{states}  \right\}  \\\\
   & \equiv & \displaystyle{ \int_\mathcal{D} \tilde{d} \, \bk \, | \bk, a \ra \la \bk , b | \qquad (a,b=1,2,3)},
\end{array}
\end{equation}
where the last equality holds {\em only} in the Hilbert spaces
spanned by the one-photon states, $\mathcal{D}$ represents the set
of \emph{detected} modes,  and we have defined the single-photon
states $ | \bk, a \ra$ as:
\begin{equation}\label{200}
 \hat{\mathcal{A}}_a^{
  \dagger}| 0 \ra = | \bk, a \ra, \qquad (a=1,2,3).
\end{equation}
From Eqs. (\ref{180}-\ref{190}) it is clear that $\hat{J}_{11},
\hat{J}_{22}$ and $\hat{J}_{33}$ form a POVM (positive operator
valued measure \cite{PeresBook}) in the space spanned by the
one-photon states.  As we saw previously, the operators
$\hat{J}_{ab}$'s allow us to introduce a $3 \times 3$ correlation
matrix operator
\begin{equation}\label{192}
 \hat{J} \equiv \left( \begin{array}{ccc}
   \displaystyle{    \hat{J}_{11}   } & \displaystyle{    \hat{J}_{21}   }
    & \displaystyle{    \hat{J}_{31}   }
   \\
   \displaystyle{    \hat{J}_{12}    } & \displaystyle{    \hat{J}_{22}   }
    & \displaystyle{    \hat{J}_{32}   }
   \\
      \displaystyle{     \hat{J}_{13}   } & \displaystyle{    \hat{J}_{23}   }
      & \displaystyle{    \hat{J}_{33}    }
  \end{array} \right).
\end{equation}
 Now, for a light beam properly
collimated around the direction $\ve^{(3)} = \vz$, the $2 \times
2$ matrix obtained by extracting the first two rows and two
columns from $\mathbb{J} \equiv \la \hat{J} \ra $, coincides with
the well known
 {\em coherency matrix} of the beam
 \cite{BornWolf}, where the bracket average $\bigl\la \cdot \bigr\ra$ is understood with respect
 to the state of the field. More generally, three independent $2 \times 2$
 matrices can be extracted from $\hat{J}$:
\begin{equation}\label{193}
\hat{J}^{(a)} = \left(
 \begin{array}{cc}
   \displaystyle{   \hat{J}_{bb}  } & \displaystyle{  \hat{J}_{cb}
   }
   \\
   \displaystyle{ \hat{J}_{bc}  } & \displaystyle{  \hat{J}_{cc}
   }
  \end{array}
  \right),
 \quad
\begin{array}{rcl}
a,b,c &\in & \{1,2,3 \},\\
 \qquad c & > &  b,  \\  c,b  & \neq & a.
 \end{array}
\end{equation}
In principle, each $\mathbb{J}^{(a)} \equiv \la \hat{J}^{(a)} \ra$
can be determined by measuring the Stokes parameters of the beam
(either classical or quantum), with a polarization analyzer whose
axis is parallel to $\ve^{(a)}$. For example, for a beam
propagating along the axis $\vz$, a set of {\em generalized}
Hermitian Stokes operators \cite{JauchBook,Aiello04c} can be
defined as
\begin{equation}\label{194}
\hat{S}_\mu \equiv \mathrm{Tr}\{ \sigma_{(\mu)} \hat{J}^{(3)} \},
\qquad (\mu = 0, \dots,3),
\end{equation}
where the $\sigma_{(\mu)}$ $(\mu=0,1,2,3)$ are the normalized
Pauli's matrices \cite{Aiello04d}. Then  from Eqs.
(\ref{193}-\ref{194}) it readily follows
\begin{equation}\label{195}
\mathbb{J}^{(3)} = \sum_{\mu = 0 }^3  s_\mu \sigma_{(\mu)},
\end{equation}
where $s_\mu \equiv \la \hat{S}_\mu \ra$.
Now we are ready to accomplish our initial task by introducing the
$2 \times 2$ effective reduced density matrix
\begin{equation}\label{210}
\rho_\mathrm{eff} \equiv \frac{\mathbb{J}^{(3)}}{\mathrm{Tr}
\mathbb{J}^{(3)}}.
\end{equation}
It is easy to check that when the set $\mathcal{D}$ of the
detected modes reduces to a single mode $\bk
\parallel \ve^{(3)}$, the definition of $\rho_\mathrm{eff}$ above
coincides with the well known polarization density matrix of a
photon \cite{DauFieldRel}. More generally, by using Eq.
(\ref{50}), it is possible to introduce an effective $3 \times 3$
reduced density matrix as $\rho \equiv \mathbb{J} / \mathrm{Tr} [
\mathbb{J}]$, which coincides with the one given by Peres {\em et
al. }\cite{Peres_et_al}.
\section*{Single-photon scattering}
Let us consider now a generic scattering process which transform
the initial single-photon density operator
$\hat{\rho}^{\mathrm{in}}$ in the output density operator
$\hat{\rho}^{\mathrm{out}}$. The most general linear
transformation between $\hat{\rho}^{\mathrm{in}}$ and
$\hat{\rho}^{\mathrm{out}}$, which leaves the density operator
Hermitian and positive semidefinite, can be written
\begin{equation}\label{n1}
\hat{\rho}^{\mathrm{out}} = \sum_{A \in \mathcal{S}} p_A
\mathcal{T}^{A } \hat{\rho}^{\mathrm{in}} \mathcal{T}^{A \dagger},
\end{equation}
where the scattering system has been represented by the ensemble
$\mathcal{S}$ of scattering matrices $\{ \mathcal{T}^{A} \}$, each
of them occurring with probability  $p_A \geq 0$. If we insert Eq.
(\ref{45}) into Eq. (\ref{n1}) we obtain
\begin{equation}\label{n2}
\begin{array}{ccl}
\hat{\rho}^{\mathrm{out}}
   & = & \displaystyle{ \sum_{\lambda, \lambda'}^{1,2} \intek \, \tek'
\rho_{\lambda \lambda'}^\mathrm{in}(\bk, \bk') \sum_{A \in
\mathcal{S}} p_A  \mathcal{T}^{A} | \bk, \lambda \ra \la \bk',
\lambda' |\mathcal{T}^{A  \dagger} },
\end{array}
\end{equation}
where
\begin{equation}\label{n4}
\begin{array}{ccl}
\displaystyle{ \sum_{A \in \mathcal{S}} p_A \mathcal{T}^{A } |
\bk, \lambda \ra \la \bk', \lambda' |\mathcal{T}^{A  \dagger} } &
= & \displaystyle{ \sum_{\theta, \theta'}^{1,2} \int \teq \, \teq'
\sum_{A \in \mathcal{S}} p_A |\bq \theta  \ra \la \bq \theta |
\mathcal{T}^{A} | \bk, \lambda \ra \la \bk', \lambda'
|\mathcal{T}^{A  \dagger} | \bq' \theta' \ra \la \bq' \theta' |}\\\\
& = & \displaystyle{ \sum_{\theta, \theta'}^{1,2} \int \teq \,
\teq' \sum_{A \in \mathcal{S}} p_A
\mathcal{T}^{A }_{\theta \lambda}(\bq,\bk) \mathcal{T}^{A \dagger
}_{ \lambda' \theta'}(\bk', \bq')
|\bq \theta  \ra \la \bq' \theta' |}.
\end{array}
\end{equation}
From the equation above, it is straightforward to see that we can
write
\begin{equation}\label{n3}
\hat{\rho}^\mathrm{out} =   \displaystyle{  \sum_{\theta,
\theta'}^{1,2} \int \teq \, \teq' \rho^\mathrm{out}_{\theta
\theta'}(\bq, \bq')
|\bq \theta  \ra \la \bq' \theta' |}
\end{equation}
where
\begin{equation}\label{n5}
 \rho^\mathrm{out}_{\theta \theta'}(\bq, \bq') =
 \sum_{A \in \mathcal{S}} p_A \sum_{\lambda, \lambda'}^{1,2}
\int \tek \, \tek'
\mathcal{T}^{A}_{\theta \lambda}(\bq,\bk) \rho_{\lambda
\lambda'}^\mathrm{in}(\bk, \bk') \mathcal{T}^{A  \dagger}_{
\lambda' \theta'}(\bk', \bq').
\end{equation}
Since $\rho^{\mathrm{t}}(\bk, \bk') = || \rho_{\lambda
\lambda'}^{\mathrm{t}}(\bk, \bk') ||$ (t = in, out) are $2 \times
2$ matrices, it is always possible to express them in the complete
Pauli basis as \cite{Aiello04d}
\begin{equation}\label{n7}
 \rho^{\mathrm{t}}(\bk, \bk') = \sum_{\mu = 0} ^3
S_\mu^\mathrm{t}(\bk, \bk') \sigma_{(\mu)},
\end{equation}
where we have introduced the \emph{two-mode Stokes parameters}
$S_\mu^\mathrm{t}(\bk, \bk') = \mathrm{Tr} \{ \sigma_{(\mu)}
\rho^{\mathrm{t}}(\bk, \bk') \}$. Since Eq. (\ref{n5}) can be
written in matrix form as
\begin{equation}\label{n7b}
 \rho^\mathrm{out}(\bq, \bq') =
 \sum_{A \in \mathcal{S}} p_A
\int \tek \, \tek'
\mathcal{T}^{A}(\bq,\bk) \rho^\mathrm{in}(\bk, \bk')
\mathcal{T}^{A  \dagger}(\bk', \bq'),
\end{equation}
 it is easy to write
\begin{equation}\label{n8}
\begin{array}{rcl}
\displaystyle{ S^\mathrm{out}_\mu (\bq, \bq')} & = &
\displaystyle{\sum_{\nu = 0} ^3 \int \tek \, \tek'
S^\mathrm{in}_\nu (\bk, \bk') \sum_{A \in \mathcal{S}} p_A
\mathrm{Tr} \left\{ \sigma_{(\mu)} \mathcal{T}^{A }(\bq,\bk)
\sigma_{(\nu)} \mathcal{T}^{A \dagger}(\bk',
\bq')\right\}}\\\\
 & \equiv &
\displaystyle{ \sum_{\nu = 0} ^3 \int \tek \, \tek'
S^\mathrm{in}_\nu (\bk, \bk') \sum_{A \in \mathcal{S}} p_A \,
m_{\mu \nu}^A}(\bq,  \bq'; \bk, \bk')\\\\
 & \equiv &
\displaystyle{ \sum_{\nu = 0} ^3 \int \tek \, \tek' M_{\mu
\nu}}(\bq,  \bq'; \bk, \bk') S^\mathrm{in}_\nu (\bk, \bk'),
\end{array}
\end{equation}
where the four-mode density Mueller matrix
\begin{equation}\label{n9}
 m_{\mu \nu}^A (\bq,  \bq'; \bk, \bk') =  \mathrm{Tr} \left\{
\sigma_{(\mu)} \mathcal{T}^{A}(\bq,\bk) \sigma_{(\nu)}
\mathcal{T}^{A \dagger} (\bk', \bq') \right\},
\end{equation}
is defined for a single ensemble realization $A$, while
\begin{equation}\label{n10}
M_{\mu \nu} (\bq,  \bq'; \bk, \bk') = \sum_{A \in \mathcal{S}} p_A
\, m_{\mu \nu}^A (\bq,  \bq'; \bk,
 \bk'),
\end{equation}
represents the ensemble-averaged four-mode density  Mueller
matrix. It is easy to see that when the input state is a
single-mode $\bk_0$ state:
\begin{equation}\label{n20}
 {\rho}^\mathrm{in}_{\lambda \lambda'} (\bk , \bk') = \rho_{\lambda \lambda'} \delta^{(3)} (\bk -
  \bk_0)\delta^{(3)} (\bk' - \bk_0),
\end{equation}
then
\begin{equation}\label{n30}
S^\mathrm{in}_{\mu} (\bk , \bk') =  s^\mathrm{in}_\mu
\delta^{(3)} (\bk -
  \bk_0)\delta^{(3)} (\bk' - \bk_0),
\end{equation}
where $\rho = || \rho_{\lambda \lambda'} ||$ and
$s^\mathrm{in}_\mu \equiv \mathrm{Tr}\{\sigma_{(\mu)} \rho \}$. In
this case, the single-mode input and output Stokes parameters
 have the same functional relation as their
classical counterparts:
\begin{equation}\label{n40}
\begin{array}{rcl}
\displaystyle{ S^\mathrm{out}_\mu (\bq_0 , \bq_0)} & = &
\displaystyle{ \sum_{\nu = 0} ^3 \int \tek \, \tek' M_{\mu
\nu}(\bq_0 , \bq_0; \bk, \bk') s^\mathrm{in}_\nu \delta^{(3)} (\bk
-
  \bk_0)\delta^{(3)} (\bk' - \bk_0)}\\\\
  & = &
\displaystyle{ \sum_{\nu = 0} ^3  M_{\mu \nu}(\bq_0 , \bq_0;
\bk_0, \bk_0) s^\mathrm{in}_\nu }\\\\
  & \equiv &
\displaystyle{ \sum_{\nu = 0} ^3  M_{\mu \nu} s^\mathrm{in}_\nu }.
\end{array}
\end{equation}

\begin{acknowledgments}
 We acknowledge support from the EU under the
IST-ATESIT contract. This project is also supported by FOM.
\end{acknowledgments}


%
%

\end{document}